\documentclass[aps,prl,preprint,groupedaddress]{revtex4-1}
\usepackage{graphicx}
\begin{document}


\title{Parity violation in $n + ^3He \rightarrow ^3H + p$ reaction: resonance approach.  }


\author{Vladimir Gudkov}
\email[]{gudkov@sc.edu}
\affiliation{Department of Physics and Astronomy, University of South Carolina, Columbia, SC, 29208}


\date{\today}

\begin{abstract}
The method based on microscopic theory of nuclear reactions has been applied for the analysis of parity violating effects in a few-body systems. Different parity violating and parity conserving  asymmetries and their dependence on neutron energy have been estimated for $n + ^3He \rightarrow ^3H + p$ reaction. The estimated effects are in a good agreement with available exact calculations.
\end{abstract}


\maketitle
\section{Introduction}

The study of parity violating (PV)  effects in low energy physics is  important for the understanding of main features of the Standard model and for the possible  search for manifestation of new physics.
During the last decades, many calculations of
different experimentally observed PV  effects in nuclear physics have been done.
However, in the last years, it became clear (see, for example \cite{Zhu:2004vw,HolsteinUSC,DesplanqueUSC,RamseyMusolf:2006dz}  and references therein) that the traditional DDH \cite{Desplanques1980} method for the calculation of PV effects cannot reliably describe the available experimental data.
This could be blamed on the ``wrong'' experimental data,
however it might be that the DDH approach is not adequate for the description of the set of precise experimental data because it is based on a number of models and assumptions. To resolve this  discrepancy,  it is desirable to increase the number of experimental data for different PV parameters in a few-body systems. The calculations of nuclear related effects for these systems can be done with  high precision, eliminating as many as possible nuclear model dependent factors involved in PV effects.
Unfortunately, currently available data of experimentally measured  PV effects in these systems are not enough to constrain all parameters required for calculations, therefore, any new potentially possible measurement is very important. Since PV effects in a few-body systems are usually very small and precise calculations of them are rather difficult, it is desirable to have a method for a reliable estimate   of possible observable parameters using available experimental data. This will give the opportunity to choose the right system and right PV observables for new experiments.

Recently, it has been proposed
to measure PV asymmetry of protons in $n + ^3He \rightarrow ^3H + p$ reaction with polarized neutrons at the Spallation Neutron Source at the Oak Ridge National Laboratory. The $^3He$ and $^4He$ systems were  subjects of intensive investigation  for a long time, and as a result,  many parameters related to reactions with neutrons and protons, as well as to excitation energy levels of these nuclei, have been measured and evaluated by a number of different groups. This rather comprehensive data  provides the opportunity to estimate  values of possible PV effects and their dependence on neutron energy  in $n + ^3He \rightarrow ^3H + p$ reaction using microscopic nuclear reaction theory approach.

\section{Description of parity violating effects}

Let us consider  the $n + ^3He \rightarrow ^3H + p$ reaction with low energy neutrons.  For neutron energy $E_n \sim 0.01\,eV$, which corresponds to a wave vector $k_n \sim 2.19\cdot 10^{-5}\, fm^{-1}$, the energy of outgoing protons and proton wave vector are $E_p = 0.764 \, MeV$ and $k_p = 0.19\, fm^{-1}$, correspondingly. Taking a characteristic $^3He$ radius as $R = 1.97\, fm$, one obtains  $(k_nR)\sim 4\cdot 10^{-4}$ and $(k_pR)\sim 0.4$. Therefore, for the initial channel,  contributions from $p$-wave neutrons to a reaction matrix (amplitude) are highly suppressed, whereas for the final channel, the amplitude with orbital momenta  of protons $l=0$ and $l=1$ have  the same order of magnitude. The contribution from $d$-wave protons is suppressed by a factor $\sim 0.025$, therefore, one can ignore $d$-waves  within the accuracy of our estimates. Assuming that  neutrons, as well as $^3He$ nuclei, may have a polarization, one shall consider four parity violating (PV) and four parity conserving (PC) angular correlations shown in Table \ref{tab_effects}, where $\vec{\sigma}$ and $\vec{I}$ are neutron and nuclear spins.
 \begin{table}[h]
 \caption{Possible parity violating  and four parity conserving  angular correlations.\label{tab_effects}}
\begin{tabular}{| c | c |}
 \hline
  PV & PC \\
  \hline
  $(\vec{\sigma}\cdot \vec{k}_p)$ & $(\vec{k}_n\cdot \vec{k}_p)$ \\
  $(\vec{\sigma}\cdot \vec{k}_n)$ & $(\vec{\sigma}\cdot [\vec{k}_n \times \vec{k}_p])$ \\
  $(\vec{I}\cdot \vec{k}_p)$ & $(\vec{\sigma}\cdot \vec{I})$ \\
  $(\vec{I}\cdot \vec{k}_n)$ & $(\vec{I}\cdot [\vec{k}_n \times \vec{k}_p])$ \\
  \hline
\end{tabular}
 \end{table}
 (It is  important to know the values of PC correlations because they usually are one of the main sources of experimental errors for PV effects, and also because they can be rather easily measured and, as a consequence, can serve as an indirect proof of the correctness of calculations of PV effects.)
We will focus on PV correlation $(\vec{\sigma}\cdot \vec{k}_p)$ which leads to PV asymmetry  $\alpha _{PV}$ of outgoing protons in a direction along to neutron polarization and opposite to it. It can be seen that the asymmetry related to the $(\vec{I}\cdot \vec{k}_p)$ correlation has exactly the same value for this reaction. For the completeness of the consideration, we will also calculate the PV effect related to differences of total cross sections $\sigma^{tot}_{\pm}$ for neutrons with opposite helicities when they propagate through  $^3He$ target, $P=(\sigma^{tot}_+-\sigma^{tot}_-)/(\sigma^{tot}_++\sigma^{tot}_-)$, which is related to the $(\vec{\sigma}\cdot \vec{k}_n)$ correlation. The corresponding difference of total cross sections for a propagation of unpolarized neutrons through polarized target, the $(\vec{I}\cdot \vec{k}_n)$ correlation, has the same value $P$ in our case. It should be noted that the $(\vec{\sigma}\cdot \vec{k}_n)$ correlation leads also to PV neutron spin precession around the direction of neutron momentum.  However, the angle of precession is very small (about $10^{-9} - 10^{-10}\; rad$ for thermal neutrons, and it can reach  the  value up to about $10^{-5}\; rad$ for $E_n \sim 0.5\;  MeV$ for the target of the size of neutron mean free path), therefore, we will not consider it here.  In addition, we will calculate the  left-right asymmetry $\alpha _{LR}$ which corresponds to the PC correlation $(\vec{\sigma}\cdot [\vec{k}_p \times \vec{k}_p])$, because it could be a source of systematic effects in the measurement of $\alpha _{PV}$.

Using standard techniques (see, for example \cite{BG_NP82}), one can represent these asymmetries in terms of matrix $\hat{R}$ which are related to reaction matrix $\hat{T}$ and to $S$-matrix as:
\begin{equation}\label{RTS}
   \hat{R}=2\pi i\hat{T}=\hat{1}-\hat{S}
\end{equation}
Then, for our case
\begin{eqnarray}\label{pv}\nonumber
 {\alpha _{PV}} &=& \frac{2}{{r}}{\mathop{\rm Re}\nolimits} [ - 3\sqrt 2  < 01|{R^1}|10 >  \cdot  < 00|{R^0}|00{ > ^*}   \\
 &+&(\sqrt 6  < 11|{R^0}|00 >  + 6 < 11|{R^1}|10 > ) < 10|{R^1}|10{ > ^*}]
 \end{eqnarray}
and
 \begin{eqnarray}\label{lr}\nonumber
 {\alpha _{LR}} &=& \frac{1}{{r}}Im[6\sqrt 3  < 10|{R^1}|11 >  \cdot  < 00|{R^0}|00{ > ^*}
  + 6\sqrt 3  < 11|{R^1}|01 >  \cdot  < 10|{R^1}|10{ > ^*}\\ \nonumber
  &+& 3(2 < 11|{R^0}|00 >  + \sqrt 6  < 11|{R^1}|10 > ) < 10|{R^1}|11{ > ^*} \\ \nonumber
  &+& 6\sqrt 2  < 00|{R^0}|11 >  \cdot  < 01|{R^1}|10{ > ^*} + 6\sqrt 3  < 10|{R^1}|01 >  \cdot  < 11|{R^1}|10{ > ^*} \\  \nonumber
  &+& \sqrt 6  < 10|{R^1}|10 > (2 < 11|{R^0}|11{ > ^*} + 3 < 11|{R^1}|11{ > ^*}) \\
  &+& 5\sqrt 6  < 11|{R^2}|11 >  \cdot  < 10|{R^1}|10{ > ^*}],
 \end{eqnarray}

where
 \begin{equation}\label{dn}
    {r} = ( |< 00|{R^0}|00 >|^2   + 3 |< 10|{R^1}|10 >|^2  ).
 \end{equation}
We use spin-channel representation, where for the matrix element $< s^{\prime} l^{\prime}|{R^J}|sl >$, $l$ and $l^{\prime}$ are orbital momenta of initial and final channels with corresponding spin-channels $s$ and $s^{\prime}$, and $J$ is the total spin of the system.
For a transmission type of observable $P$, one obtains:
\begin{eqnarray}\label{P}\nonumber
 P =  &-& \frac{{{\mathop{\rm Re}\nolimits} [ < 00|{R^0}|11 >  +  < 11|{R^0}|00 > ]}}{{{\mathop{\rm Re}\nolimits} [ < 00|{R^0}|00 >  + 3 < 10|{R^1}|10 > ]}} \\ \nonumber
  &+& \sqrt 3 \frac{{{\mathop{\rm Re}\nolimits} [ < 10|{R^1}|01 >  +  < 01|{R^1}|10 > ]}}{{{\mathop{\rm Re}\nolimits} [ < 00|{R^0}|00 >  + 3 < 10|{R^1}|10 > ]}} \\ \nonumber
  &+& \sqrt 6 \frac{{{\mathop{\rm Re}\nolimits} [ < 10|{R^1}|11 >  +  < 11|{R^1}|10 > ]}}{{{\mathop{\rm Re}\nolimits} [ < 00|{R^0}|00 >  + 3 < 10|{R^1}|10 > ]}} .
 \end{eqnarray}

\begin{figure}
\includegraphics{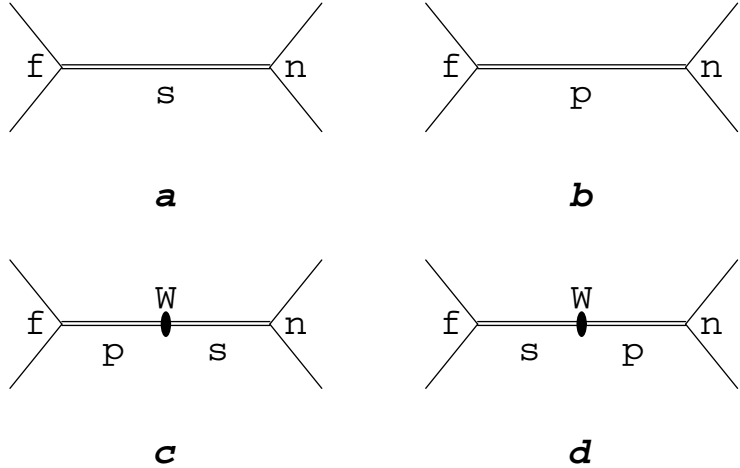}
\caption{Parity conserving   (diagrams a and b) and parity violating (diagrams c and d) matrix elements of $\hat{R}$. Symbols $s$ and $p$ corresponds to $s$- and $p$-wave neutrons for the initial (neutron) channel, and indicate parity of the final channel. \label{fig1}}
 \end{figure}

Calculations of  matrix elements $< s^{\prime} l^{\prime}|{R^J}|sl >$ for parity
violating effects in nuclear reactions have been done \cite{BG_NP82} using distorted wave Born approximation  in microscopic theory of nuclear reactions \cite{MW}.  They lead to
the symmetry violating amplitudes induced by parity violating potential $W$
\begin{equation}
R^{fi}_{PV} = 2\pi i<{\Psi^-_f}|W|{\Psi^+_i}>,
\end{equation}
 where $\Psi^{\pm}_{i,f}$ are the eigenfunctions of the nuclear P-invariant
 Hamiltonian with  the  appropriate boundary conditions \cite{MW}:
\begin{equation}
 \Psi^{\pm}_{i,f}=\sum_k a^\pm_{k(i,f)}(E)\; \phi_k + \sum_m\int
b^{\pm}_{m(i,f)}(E,E')\; \chi^{\pm}_m(E')\; dE'.
 \label{eq:wf}
\end{equation}
Here, $\phi_k$ is the wave function of  the $k^{th}$
  resonance and $\chi^{\pm}_m(E)$ is the potential
 scattering wave function in the channel $m$.
The coefficient
\begin{equation}
 a^\pm_{k(i,f)}(E)={\exp{(\pm i\delta_{i,f})}\over {(2\pi)^{1\over 2}}}{{(\Gamma^{i,f}_k)^{1\over 2}}\over {E-E_k\pm{i\over
   2}\Gamma_k}}
\end{equation}
describes  nuclear resonances contributions, and the
coefficient $b^{\pm}_{m(i,f)}(E,E')$ describes potential
scattering and interactions between the continuous spectrum and
 resonances. (Here, $E_k$, $\Gamma_k$, and $\Gamma^i_k$ are
the energy, the total width, and the partial width in the channel
$i$ of the $k$-th  resonance, $E$ is the neutron
energy, and $\delta_i$ is the potential scattering phase in the
channel $i$; $(\Gamma^i_k)^{1\over 2} = (2\pi )^{1\over 2}
 <{\chi_i(E)}|V|{\phi_k}>$,
 where $V$ is a residual interaction operator.)
 As it was shown in \cite{BG_NP82} for nuclei with rather large atomic numbers, the resonance contribution  is  dominant. Then, for the simplest case with only two resonances with opposite parities,  the expressions for matrix element  $\hat{R}$ for neutron-proton reaction  with parity violation is:
 \begin{equation}
< s^{\prime} l^{\prime}|{R^J}|sl > = -
{{iw(\Gamma^n_l(s)\Gamma^p_{l^{\prime}}(s^{\prime}))^{1\over
2}}\over{(E-E_l+i\Gamma_l/2)(E-E_{l^{\prime}}+i\Gamma_{l^{\prime}}/2)}}{\it
e}^{i(\delta^n_l + \delta^p_{l^{\prime}})}, \label{eq:pv}
\end{equation}
and with conservation of parity (one resonance contribution) is:
\begin{equation}
< s^{\prime} l^{\prime}|{R^J}|sl > =
{{i(\Gamma^n_l(s)\Gamma^p_{l^{\prime}}(s^{\prime}))^{1\over
2}}\over{(E-E_l+i\Gamma_l/2)}}{\it
e}^{i(\delta^n_l + \delta^p_{l^{\prime}})}, \label{eq:pc}
\end{equation}
where $w=-\int \phi_l W \phi_{l^{\prime}}d\tau$ is parity violating nuclear matrix element mixing parities of two resonances. The above $\hat{R}$ matrix elements could be represented by diagrams  shown on Fig.(\ref{fig1}). Thus, PV asymmetry ${\alpha _{PV}}$ is proportional to the real part the product of $\hat{R}$ matrices  presented by diagrams $a$ and $c$, and PC asymmetry ${\alpha _{LR}}$ is proportional to the imaginary part of the product of $\hat{R}$ matrices  presented by diagrams $a$ and $b$ (see for details \cite{BG_NP82}).

This technique has been proven to work very well for calculation of nuclear PV effects for intermediate and heavy nuclei. Assuming the dominant resonance contribution to PV effects for $n + ^3He \rightarrow ^3H + p$ reaction, we will apply  this approach  to estimate  characteristic values of PV  effects  using parametrization of PV effects in terms of known  resonance structure of the system. Fortunately,  the detailed structure of resonances ($^4He$ levels) \cite{Tilley:1992} and low energy neutron scattering parameters \cite{Mughabghab} are well known for this reaction from numerous experiments.

To estimate PV and PC asymmetries in  $n + ^3He \rightarrow ^3H + p$ reactions using the described above formalism, we will take into account all known resonances \cite{Tilley:1992,Mughabghab} which result in multi-resonance representation for $\hat{R}$ matrix elements. From the selection rules for angular momenta (see Eqs. (\ref{pv}), (\ref{lr}), (\ref{P}) and general expressions in \cite{BG_NP82}), one can see that  for low energy neutrons only resonances with the total spin of $J=0,1$  contribute to PV asymmetries of the interest. However,  for the left-right PC asymmetry, we have to consider $J=0,1,2$. Thus, for PV effects we consider contributions from nine low energy resonances \cite{Tilley:1992,Mughabghab} (see Table (\ref{tab_res1}) ): one resonance with total angular momentum and parity $J^{\pi}=0^+$, three with $J^{\pi}=0^-$, four with $J^{\pi}=1^-$, and one with $J^{\pi}=1^+$. For further calculations, we assume that all weak matrix elements, which mix resonances with opposite parities,   have the same values and are described by a phenomenological formula \cite{BG_NP82} $w=2\cdot10^{-4}eV\sqrt{\bar{D}(eV)}$ (where  $\bar{D}$ is an average energy level spacing). This formula is in good agreement with other statistical nuclear model estimates \cite{Kadmensky:1983,Bunakov:1989,Johnson:1991} of nuclear weak matrix elements for medium and heavy nuclei. The extrapolation of this formula to the region of one-particle nuclear excitation  leads to correct  value for weak nucleon-nucleon interaction. Therefore, one can use this approximation for  rough estimates  of average values of weak matrix elements in few-body systems. This leads to the value of weak matrix element $w=0.5\;eV$ (with $\bar{D}\simeq 6\; MeV$),  which is rather close to the typical value of one particle weak matrix element. One can see from Eqs.(\ref{eq:pv}) and (\ref{eq:pc}) that the expressions for PV and PC $\hat{R}$ matrices depend not on  neutron and proton partial widths   but on their  amplitudes,  the values of which depend on particular spin-channels. Since we know only partial widths, we have to make  assumptions about values of amplitudes of partial widths for a specific spin-channel and about their signs (phases).   This leads to another uncertainty in our estimation in addition to the given above assumption about weak matrix elements.   To treat the spin-channel dependence of partial width amplitudes, we assume that  partial widths for each spin-channel are equal to each other.  This gives us an average factor of uncertainly of about $2$. The signs of  width amplitudes, as well as the signs of weak matrix elements $w$, are left undetermined (random). This also can lead to a factor of uncertainly of $2$ or $3$. Therefore, one can see that the uncertainly of our multi-resonance calculations is about of one order of magnitude.

\section{Discussion of Result of Calculations }

Taking into account the considerations given above and using resonance parameters \cite{Tilley:1992,Mughabghab} of the Table (\ref{tab_res1}), one can estimate the PV asymmetry for thermal neutrons   as
\begin{equation}\label{pv1}
    {\alpha _{PV}}=-(1-4)\cdot 10^{-7}.
\end{equation}
The set of resonance parameters of the Table (\ref{tab_res2}) results in a slightly lager   PV asymmetry
\begin{equation}\label{pv2}
    {\alpha _{PV}}=-(4-8)\cdot 10^{-7}.
\end{equation}
The difference of these two sets  is related to the discrepancy between \cite{Mughabghab} and \cite{Tilley:1992} for resonance parameters for the first positive resonance ($E_n=0.430\; MeV$).

The left-right asymmetry at thermal energy is less sensitive to the parameters of this first resonance and has about the same value for these two sets of resonance parameters
\begin{equation}\label{lr1}
    {\alpha _{LR}}=-(2-8)\cdot 10^{-4}.
\end{equation}
It should be noted that for the calculation of the left-right asymmetry, we  add first three  $2^-$ resonances \cite{Tilley:1992} which do not contribute to PV effects for low energy neutrons.

\begin{figure}
\includegraphics{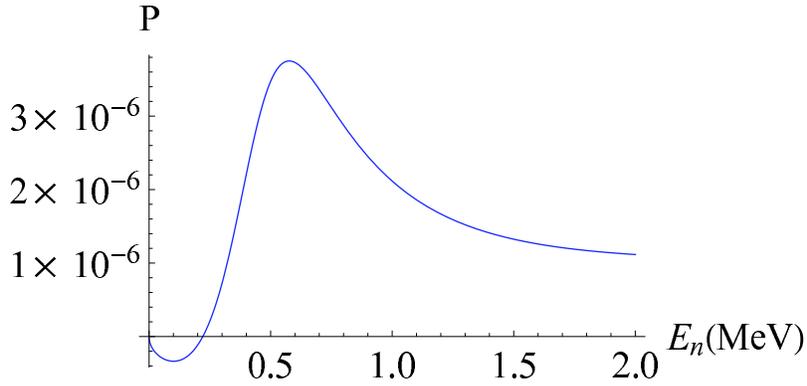}
\caption{(Color online) Resonance enhancement of of the $P$ parameter. \label{Pplot}}
 \end{figure}

\begin{figure}
\includegraphics{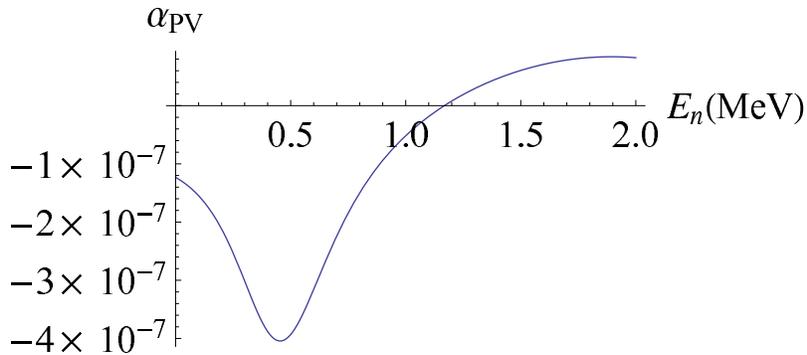}
\caption{(Color online) Resonance enhancement of of the $\alpha _{PV}$ asymmetry (for the first set of parameters). \label{a1plot}}
 \end{figure}

The value of PV in neutron transmission for the first choice of parameters is
\begin{equation}\label{tr1}
    P=-(2-4)\cdot 10^{-10},
\end{equation}
 and for the second choice is
\begin{equation}\label{tr2}
    P=-(0.8-1.6)\cdot 10^{-10}.
\end{equation}
 One can see that the parameter $P$ is very small for neutrons with thermal energy, but it is essentially enhanced in a few-$MeV$ region  (see Fig.(\ref{Pplot}). The asymmetry ${\alpha _{PV}}$ also shows a resonance behavior but its enhancement is not very large (see Fig.(\ref{a1plot})).

To show  contributions of each resonance to PV asymmetry $\alpha _{PV}$ and to transmission parameter $P$, we normalized contributions from each resonance in terms of relative intensity  to the most strong one, which is taken as $100\%$ (see last two columns in  Tables (\ref{tab_res1}) and (\ref{tab_res2})).  Some resonances contribute through two different spin-channels $s=0$ and $s=1$. In those cases, the contributions from two spin-channels can be either with the same sign or with the opposite sign, depending on unknown phases of amplitudes of partial widths and weak matrix elements (see, for example, resonance at $3.062\; MeV$ in Table (\ref{tab_res1})). As can be seen from these tables,   different resonances contribute essentially differently to the value of PV violating effects (this is also correct for parity conserving asymmetries). 
Moreover,  different sets of resonance parameters can change  weights of the resonances for a particular asymmetry. For example, the lowest $0^-$-resonance contribution to  the asymmetry ${\alpha _{PV}}$ appears to be 3\% using the parameters of Table (\ref{tab_res1}), while it would be dominant one using the parameters of Table (\ref{tab_res2}). This is related to the fact that for the set of Table (\ref{tab_res1}) the contribution of the $0^-$-resonance to the ${\alpha _{PV}}$ is suppressed by a factor of about 40 due to destructive interference between parity conserving and parity violating amplitudes.
Therefore, the readability of this method can be essentially improved by increasing the accuracy in measurements of parameters of the most ``important'' resonances.

It should be noted that the estimated value of the PV asymmetry $\alpha _{PV}$ at thermal energy (see Eqs. (\ref{pv1}) and (\ref{pv2})) is surprisingly in very good agreement with exact calculations for zero energy neutrons \cite{Viviani:2010qt}. This could be considered as an additional argument for reliability of the suggested resonance approach. Also, matching the estimated value of the observable parameter with exact calculations at low energy gives us the opportunity to predict PV effects in a wide range of neutron energies.

\begin{table}
 \caption{Resonance parameters (Set 1). Here $E_r$ is a resonance energy; $T$ and $J^{\pi}$  are resonance isospin and the total resonance spin with parity; $\Gamma$ and $\Gamma_p$ are total and proton widths; $\Gamma_n$, $\Gamma^0_n$ and  $l$ are neutron width, reduced width, and angular momentum, correspondingly;  $\alpha _{PV}$ (\% ) and $P$ (\% ) are normalized contribution of the resonance to  $\alpha _{PV}$ and $P$, correspondingly. \label{tab_res1}}
\begin{tabular}{| c | c | c | c | c | c | c | c | c | c |}
 \hline
  $E_r (MeV)$&$\;J^{\pi}\;$&$\;\; l\;\; $&$\;\;T\;\;$& $\Gamma_n(MeV)$ & $\Gamma^0_n(eV)$ &  $\Gamma_p(MeV)$ &  $\Gamma(MeV)$ & $\alpha _{PV}$ (\% ) & $P$ (\% )\\
 \hline
-0.211 & 0+ & 0 & 0 &   & 954.4 & 1.153 & 1.153 &  & \\
0.430 & 0- & 1 & 0 & 0.48  &   & 0.05 & 0.53 & 3.1 & 100\\
3.062 & 1- & 1 & 1 & 2.76  &   & 3.44 & 6.20 & 100$\pm$26 & 2$\pm$1\\
3.672 & 1- & 1 & 0 & 2.87  &   & 3.08 & 6.10 & 75$\pm$24 & 1$\pm$1\\
4.702 & 0- & 1 & 1 & 3.85  &   & 4.12 & 7.97 & 20 & 3\\
5.372 & 1- & 1 & 1 & 6.14  &   & 6.52 & 12.66 & 79$\pm$18 & 1\\
7.732 & 1+ & 0 & 0 & 4.66  &   & 4.725 & 9.89 &  & \\
7.792 & 1- & 1 & 0 & 0.08  &   & 0.07 & 3.92 & 2$\pm$1 & 0 \\
8.062 & 0- & 1 & 0 & 0.01  &   & 0.01 & 4.89 & 14 & 0\\
  \hline
\end{tabular}
 \end{table}

\begin{table}
 \caption{Resonance parameters (Set 2). Here $E_r$ is a resonance energy; $T$ and $J^{\pi}$  are resonance isospin and the total resonance spin with parity; $\Gamma$ and $\Gamma_p$ are total and proton widths; $\Gamma_n$, $\Gamma^0_n$ and  $l$ are neutron width, reduced width, and angular momentum, correspondingly;  $\alpha _{PV}$ (\% ) and $P$ (\% ) are normalized contribution of the resonance to  $\alpha _{PV}$ and $P$, correspondingly.\label{tab_res2}}
\begin{tabular}{| c | c | c | c | c | c | c | c | c | c |}
 \hline
  $E_r (MeV)$&$\;J^{\pi}\;$&$\;\; l\;\; $&$\;\;T\;\;$& $\Gamma_n(MeV)$ & $\Gamma^0_n(eV)$ &  $\Gamma_p(MeV)$ &  $\Gamma(MeV)$ & $\alpha _{PV}$ (\% ) & $P$ (\% )\\
 \hline
-0.211 & 0+ & 0 & 0 &   & 954.4 & 1.153 & 1.153 &   & \\
0.430 & 0- & 1 & 0 & 0.20  &   & 0.640 & 0.84 & 100 & 100\\
3.062 & 1- & 1 & 1 & 2.76  &   & 3.44 & 6.20 & 82$\pm$27 & 4$\pm$3\\
3.672 & 1- & 1 & 0 & 2.87  &   & 3.08 & 6.10 & 62$\pm$20 & 3$\pm$2\\
4.702 & 0- & 1 & 1 & 3.85  &   & 4.12 & 7.97 & 16 & 8\\
5.372 & 1- & 1 & 1 & 6.14  &   & 6.52 & 12.66 & 65$\pm$15 & 2$\pm$1.5\\
7.732 & 1+ & 0 & 0 & 4.66  &   & 4.725 & 9.89 &  & \\
7.792 & 1- & 1 & 0 & 0.08  &   & 0.07 & 3.92 & 2$\pm$1 & 0\\
8.062 & 0- & 1 & 0 & 0.01  &   & 0.01 & 4.89 & 1 & 0.5\\
  \hline
\end{tabular}
 \end{table}

\begin{acknowledgments}
I am grateful Dr. J. D. Bowman for fruitful discussions.
This work was supported by the DOE grants no. DE-FG02-09ER41621.
\end{acknowledgments}

\bibliography{n3He}

\providecommand{\noopsort}[1]{}\providecommand{\singleletter}[1]{#1}%
\begin{thebibliography}{13}%
\makeatletter
\providecommand \@ifxundefined [1]{%
 \@ifx{#1\undefined}
}%
\providecommand \@ifnum [1]{%
 \ifnum #1\expandafter \@firstoftwo
 \else \expandafter \@secondoftwo
 \fi
}%
\providecommand \@ifx [1]{%
 \ifx #1\expandafter \@firstoftwo
 \else \expandafter \@secondoftwo
 \fi
}%
\providecommand \natexlab [1]{#1}%
\providecommand \enquote  [1]{``#1''}%
\providecommand \bibnamefont  [1]{#1}%
\providecommand \bibfnamefont [1]{#1}%
\providecommand \citenamefont [1]{#1}%
\providecommand \href@noop [0]{\@secondoftwo}%
\providecommand \href [0]{\begingroup \@sanitize@url \@href}%
\providecommand \@href[1]{\@@startlink{#1}\@@href}%
\providecommand \@@href[1]{\endgroup#1\@@endlink}%
\providecommand \@sanitize@url [0]{\catcode `\\12\catcode `\$12\catcode
  `\&12\catcode `\#12\catcode `\^12\catcode `\_12\catcode `\%12\relax}%
\providecommand \@@startlink[1]{}%
\providecommand \@@endlink[0]{}%
\providecommand \url  [0]{\begingroup\@sanitize@url \@url }%
\providecommand \@url [1]{\endgroup\@href {#1}{\urlprefix }}%
\providecommand \urlprefix  [0]{URL }%
\providecommand \Eprint [0]{\href }%
\@ifxundefined \urlstyle {%
  \providecommand \doi  [0]{\begingroup \@sanitize@url \@doi}%
  \providecommand \@doi [1]{\endgroup \@@startlink {\doibase
  #1}doi:\discretionary {}{}{}#1\@@endlink }%
}{%
  \providecommand \doi  [0]{doi:\discretionary{}{}{}\begingroup
  \urlstyle{rm}\Url }%
}%
\providecommand \doibase [0]{http://dx.doi.org/}%
\providecommand \Doi [0]{\begingroup \@sanitize@url \@Doi }%
\providecommand \@Doi  [1]{\endgroup\@@startlink{\doibase#1}\@@Doi}%
\providecommand \@@Doi [1]{#1\@@endlink}%
\providecommand \selectlanguage [0]{\@gobble}%
\providecommand \bibinfo  [0]{\@secondoftwo}%
\providecommand \bibfield  [0]{\@secondoftwo}%
\providecommand \translation [1]{[#1]}%
\providecommand \BibitemOpen [0]{}%
\providecommand \bibitemStop [0]{}%
\providecommand \bibitemNoStop [0]{.\EOS\space}%
\providecommand \EOS [0]{\spacefactor3000\relax}%
\providecommand \BibitemShut  [1]{\csname bibitem#1\endcsname}%
\bibitem [{\citenamefont {Zhu}\ \emph {et~al.}(2005)\citenamefont {Zhu},
  \citenamefont {Maekawa}, \citenamefont {Holstein}, \citenamefont
  {Ramsey-Musolf},\ and\ \citenamefont {van Kolck}}]{Zhu:2004vw}%
  \BibitemOpen
  \bibfield  {author} {\bibinfo {author} {\bibfnamefont {S.-L.}\ \bibnamefont
  {Zhu}}, \bibinfo {author} {\bibfnamefont {C.~M.}\ \bibnamefont {Maekawa}},
  \bibinfo {author} {\bibfnamefont {B.~R.}\ \bibnamefont {Holstein}}, \bibinfo
  {author} {\bibfnamefont {M.~J.}\ \bibnamefont {Ramsey-Musolf}}, \ and\
  \bibinfo {author} {\bibfnamefont {U.}~\bibnamefont {van Kolck}},\ }\Doi
  {10.1016/j.nuclphysa.2004.10.032} {\bibfield  {journal} {\bibinfo  {journal}
  {Nucl. Phys.},\ }\textbf {\bibinfo {volume} {A748}},\ \bibinfo {pages} {435}
  (\bibinfo {year} {2005})}\BibitemShut {NoStop}%
\bibitem [{\citenamefont {Holstein}(2005)}]{HolsteinUSC}%
  \BibitemOpen
  \bibfield  {author} {\bibinfo {author} {\bibfnamefont {B.}~\bibnamefont
  {Holstein}},\ }\href@noop {} {\enquote {\bibinfo {title} {Neutrons and
  hadronic parity violation},}\ } (\bibinfo {year} {2005}),\ \bibinfo {note}
  {proc. of Int. Workshop on Theoretical Problems in Fundamental Neutron
  Physics, October 14-15, 2005, Columbia, SC},\ \Eprint
  {http://arxiv.org/abs/http://www.physics.sc.edu/TPFNP/Talks/Program.html}
  {http://www.physics.sc.edu/TPFNP/Talks/Program.html} \BibitemShut {NoStop}%
\bibitem [{\citenamefont {Desplanque}(2005)}]{DesplanqueUSC}%
  \BibitemOpen
  \bibfield  {author} {\bibinfo {author} {\bibfnamefont {B.}~\bibnamefont
  {Desplanque}},\ }\href@noop {} {\enquote {\bibinfo {title} {Weak couplings: a
  few remarks},}\ } (\bibinfo {year} {2005}),\ \bibinfo {note} {proc. of Int.
  Workshop on Theoretical Problems in Fundamental Neutron Physics, October
  14-15, 2005, Columbia, SC},\ \Eprint
  {http://arxiv.org/abs/http://www.physics.sc.edu/TPFNP/Talks/Program.html}
  {http://www.physics.sc.edu/TPFNP/Talks/Program.html} \BibitemShut {NoStop}%
\bibitem [{\citenamefont {Ramsey-Musolf}\ and\ \citenamefont
  {Page}(2006)}]{RamseyMusolf:2006dz}%
  \BibitemOpen
  \bibfield  {author} {\bibinfo {author} {\bibfnamefont {M.~J.}\ \bibnamefont
  {Ramsey-Musolf}}\ and\ \bibinfo {author} {\bibfnamefont {S.~A.}\ \bibnamefont
  {Page}},\ }\Doi {10.1146/annurev.nucl.54.070103.181255} {\bibfield  {journal}
  {\bibinfo  {journal} {Ann. Rev. Nucl. Part. Sci.},\ }\textbf {\bibinfo
  {volume} {56}},\ \bibinfo {pages} {1} (\bibinfo {year} {2006})},\ \Eprint
  {http://arxiv.org/abs/hep-ph/0601127} {arXiv:hep-ph/0601127} \BibitemShut
  {NoStop}%
\bibitem [{\citenamefont {Desplanques}\ \emph {et~al.}(1980)\citenamefont
  {Desplanques}, \citenamefont {Donoghue},\ and\ \citenamefont
  {Holstein}}]{Desplanques1980}%
  \BibitemOpen
  \bibfield  {author} {\bibinfo {author} {\bibfnamefont {B.}~\bibnamefont
  {Desplanques}}, \bibinfo {author} {\bibfnamefont {J.~F.}\ \bibnamefont
  {Donoghue}}, \ and\ \bibinfo {author} {\bibfnamefont {B.~R.}\ \bibnamefont
  {Holstein}},\ }\Doi {DOI: 10.1016/0003-4916(80)90217-1} {\bibfield  {journal}
  {\bibinfo  {journal} {Annals of Physics},\ }\textbf {\bibinfo {volume}
  {124}},\ \bibinfo {pages} {449 } (\bibinfo {year} {1980})},\ ISSN \bibinfo
  {issn} {0003-4916}\BibitemShut {NoStop}%
\bibitem [{\citenamefont {Bunakov}\ and\ \citenamefont
  {Gudkov}(1983)}]{BG_NP82}%
  \BibitemOpen
  \bibfield  {author} {\bibinfo {author} {\bibfnamefont {V.~E.}\ \bibnamefont
  {Bunakov}}\ and\ \bibinfo {author} {\bibfnamefont {V.~P.}\ \bibnamefont
  {Gudkov}},\ }\Doi {10.1016/0375-9474(83)90338-X} {\bibfield  {journal}
  {\bibinfo  {journal} {Nucl. Phys.},\ }\textbf {\bibinfo {volume} {A401}},\
  \bibinfo {pages} {93} (\bibinfo {year} {1983})}\BibitemShut {NoStop}%
\bibitem [{\citenamefont {Mahaux}\ and\ \citenamefont
  {Weidenmuller}(1969)}]{MW}%
  \BibitemOpen
  \bibfield  {author} {\bibinfo {author} {\bibfnamefont {C.}~\bibnamefont
  {Mahaux}}\ and\ \bibinfo {author} {\bibfnamefont {H.~A.}\ \bibnamefont
  {Weidenmuller}},\ }\href@noop {} {\emph {\bibinfo {title} {Shell-model
  approach to nuclear reactions}}}\ (\bibinfo  {publisher} {North-Holland Pub.
  Co., Amsterdam, London},\ \bibinfo {year} {1969})\BibitemShut {NoStop}%
\bibitem [{\citenamefont {Tilley}\ \emph {et~al.}(1992)\citenamefont {Tilley},
  \citenamefont {Weller},\ and\ \citenamefont {Hale}}]{Tilley:1992}%
  \BibitemOpen
  \bibfield  {author} {\bibinfo {author} {\bibfnamefont {D.~R.}\ \bibnamefont
  {Tilley}}, \bibinfo {author} {\bibfnamefont {H.~R.}\ \bibnamefont {Weller}},
  \ and\ \bibinfo {author} {\bibfnamefont {G.~M.}\ \bibnamefont {Hale}},\
  }\href@noop {} {\bibfield  {journal} {\bibinfo  {journal} {Nucl. Phys.},\
  }\textbf {\bibinfo {volume} {A541}},\ \bibinfo {pages} {1} (\bibinfo {year}
  {1992})}\BibitemShut {NoStop}%
\bibitem [{\citenamefont {Mughabghab}(2006)}]{Mughabghab}%
  \BibitemOpen
  \bibfield  {author} {\bibinfo {author} {\bibfnamefont {S.~F.}\ \bibnamefont
  {Mughabghab}},\ }\href@noop {} {\emph {\bibinfo {title} {Atlas of Neutron
  Resonances: Resonance Parameters and Thermal Cross Sections. Z=1-100;
  electronic version}}}\ (\bibinfo  {publisher} {Elsevier},\ \bibinfo {address}
  {San Diego, CA},\ \bibinfo {year} {2006})\BibitemShut {NoStop}%
\bibitem [{\citenamefont {Kadmensky}\ \emph {et~al.}(1983)\citenamefont
  {Kadmensky}, \citenamefont {Markushev},\ and\ \citenamefont
  {Furman}}]{Kadmensky:1983}%
  \BibitemOpen
  \bibfield  {author} {\bibinfo {author} {\bibfnamefont {S.~G.}\ \bibnamefont
  {Kadmensky}}, \bibinfo {author} {\bibfnamefont {V.~P.}\ \bibnamefont
  {Markushev}}, \ and\ \bibinfo {author} {\bibfnamefont {V.~I.}\ \bibnamefont
  {Furman}},\ }\href@noop {} {\bibfield  {journal} {\bibinfo  {journal} {Sov.
  J. Nucl. Phys.},\ }\textbf {\bibinfo {volume} {37}},\ \bibinfo {pages} {345}
  (\bibinfo {year} {1983})}\BibitemShut {NoStop}%
\bibitem [{\citenamefont {Bunakov}\ \emph {et~al.}(1989)\citenamefont
  {Bunakov}, \citenamefont {Gudkov}, \citenamefont {Kadmensky}, \citenamefont
  {Lomachenkov},\ and\ \citenamefont {Furman}}]{Bunakov:1989}%
  \BibitemOpen
  \bibfield  {author} {\bibinfo {author} {\bibfnamefont {V.~E.}\ \bibnamefont
  {Bunakov}}, \bibinfo {author} {\bibfnamefont {V.~P.}\ \bibnamefont {Gudkov}},
  \bibinfo {author} {\bibfnamefont {S.~G.}\ \bibnamefont {Kadmensky}}, \bibinfo
  {author} {\bibfnamefont {I.~A.}\ \bibnamefont {Lomachenkov}}, \ and\ \bibinfo
  {author} {\bibfnamefont {V.~I.}\ \bibnamefont {Furman}},\ }\href@noop {}
  {\bibfield  {journal} {\bibinfo  {journal} {Sov. J. Nucl. Phys.},\ }\textbf
  {\bibinfo {volume} {49}},\ \bibinfo {pages} {613} (\bibinfo {year}
  {1989})}\BibitemShut {NoStop}%
\bibitem [{\citenamefont {Johnson}\ \emph {et~al.}(1991)\citenamefont
  {Johnson}, \citenamefont {Bowman},\ and\ \citenamefont {Yoo}}]{Johnson:1991}%
  \BibitemOpen
  \bibfield  {author} {\bibinfo {author} {\bibfnamefont {M.~B.}\ \bibnamefont
  {Johnson}}, \bibinfo {author} {\bibfnamefont {J.~D.}\ \bibnamefont {Bowman}},
  \ and\ \bibinfo {author} {\bibfnamefont {S.~H.}\ \bibnamefont {Yoo}},\ }\Doi
  {10.1103/PhysRevLett.67.310} {\bibfield  {journal} {\bibinfo  {journal}
  {Phys. Rev. Lett.},\ }\textbf {\bibinfo {volume} {67}},\ \bibinfo {pages}
  {310} (\bibinfo {year} {1991})}\BibitemShut {NoStop}%
\bibitem [{\citenamefont {Viviani}\ \emph {et~al.}(2010)\citenamefont
  {Viviani}, \citenamefont {Schiavilla}, \citenamefont {Girlanda},
  \citenamefont {Kievsky},\ and\ \citenamefont {Marcucci}}]{Viviani:2010qt}%
  \BibitemOpen
  \bibfield  {author} {\bibinfo {author} {\bibfnamefont {M.}~\bibnamefont
  {Viviani}}, \bibinfo {author} {\bibfnamefont {R.}~\bibnamefont {Schiavilla}},
  \bibinfo {author} {\bibfnamefont {L.}~\bibnamefont {Girlanda}}, \bibinfo
  {author} {\bibfnamefont {A.}~\bibnamefont {Kievsky}}, \ and\ \bibinfo
  {author} {\bibfnamefont {L.~E.}\ \bibnamefont {Marcucci}},\ }\Doi
  {10.1103/PhysRevC.82.044001} {\bibfield  {journal} {\bibinfo  {journal}
  {Phys. Rev.},\ }\textbf {\bibinfo {volume} {C82}},\ \bibinfo {pages} {044001}
  (\bibinfo {year} {2010})}\BibitemShut {NoStop}%
\end{thebibliography}%

\end{document}